\documentclass[11pt,a4paper]{article}
\pdfoutput=1
\usepackage{jcappub}
\graphicspath{{images/}}
\newcommand{\myat}{{\fontfamily{pag}\selectfont @}}
\title{\boldmath 
Probing a cosmic axion-like particle background within the jets of active galactic nuclei
}

\author[1]{Ahmed Ayad\note{Corresponding author.}}
\author{and Geoff Beck}
\affiliation{School of Physics, University of the Witwatersrand, Private Bag 3, WITS-2050, Johannesburg, South Africa}

\emailAdd{ahmed\myat aims.edu.gh}
\emailAdd{Geoffrey.Beck\myat wits.ac.za }

\abstract{Axions or more generally axion-like particles (ALPs) are pseudo-scalar particles predicted by many extensions of the Standard Model of particle physics (SM) and considered as viable candidates for dark matter (DM) in the universe. If they really exist in nature, they are expected to couple with photons in the presence of an external electromagnetic field through a form of the Primakoff effect. In addition, many string theory models of the early universe motivate the existence of a homogeneous Cosmic ALP Background (CAB) with $0.1 \textup{--} 1$ keV energies analogous to the Cosmic Microwave Background (CMB), arising via the decay of string theory moduli in the very early universe. The coupling between the CAB ALPs traveling in cosmic magnetic fields and photons allows ALPs to oscillate into photons and vice versa. In this work, we test the CAB model that is put forward to explain the soft X-ray excess in the Coma cluster due to CAB ALPs conversion into photons using the M87 jet environment. Then we demonstrate the potential of the active galactic nuclei (AGNs) jet environment to probe low-mass ALP models, and to potentially constrain the model proposed to better explain the Coma cluster soft X-ray excess. We find that the overall X-ray emission for the M87 AGN requires an ALP-photon coupling $g_{a\gamma}$ in the range of $\sim 7.50 \times 10^{-15} \textup{--} 6.56 \times 10^{-14} \; \text{GeV}^{-1}$ for ALP masses $m_a \lesssim 10^{-13} \; \text{eV}$ as long as the M87 jet is misaligned by less than about 20 degrees from the line of sight. These values are up to an order of magnitude smaller than the current best fit value on $g_{a\gamma} \sim 2 \times 10^{-13} \; \text{GeV}^{-1}$ obtained in soft X-ray excess CAB model for the Coma cluster. Our results cast doubt on the current limits of the largest allowed value of $g_{a\gamma}$ and suggest a new constraint that $g_{a\gamma} \lesssim 6.56 \times 10^{-14} \; \text{GeV}^{-1}$ when a CAB is assumed.}

\arxivnumber{1911.10078}
\begin{document}
\maketitle
\flushbottom

\section{Introduction}

An outstanding result of modern cosmology is that only a small fraction of the total matter content of the universe is made of baryonic matter, while the vast majority is constituted by dark matter (DM) \cite{komatsu2011seven}. However, the nature of such a component is still unknown. One theoretically well motivated approach would be to consider light scalar candidates of DM such as axions and axion-like particles (ALPs). Axions \cite{peccei1977cp, weinberg1978new} are pseudo-Nambu-Goldstone bosons that appear after the spontaneous breaking of the Peccei-Quinn symmetry introduced to solve the CP-violation problem of the strong interaction, which represents one of the serious problems in the standard model of particle physics (SM), for a review see reference \cite{peccei2008strong}. The theory, together with observational and experimental bounds, predicts that such axions are very light and weakly interacting with the SM particles, see reference \cite{asztalos2006searches}.  For these reasons, they are suggested to be suitable candidates for the DM content of the universe \cite{preskill1983cosmology, abbott1983cosmological, dine1983not}. The observation of a light axion would indeed solve the strong CP-problem and at least would participate in improving our understanding of the origin of the component of the DM in the universe. Furthermore, ALPs are pseudo-scalar particles predicted in many theoretically well-motivated extensions to the SM \cite{arvanitaki2010string, cicoli2012type, anselm1982second}. Axions and ALPs are characterized by their coupling with two photons. While the coupling parameter for axions is related to the axion mass, there is no direct relation between the coupling parameter and the masses of ALPs. Nevertheless, it is expected that ALPs share the same phenomenology of axions. Therefore, it is argued that axions or now more generally ALPs \cite{arias2012wispy, ringwald2012exploring, marsh2016axion} are highly viable candidates for DM in the universe.

If ALPs really exist in nature, they are expected to couple with photons in the presence of an external electric or magnetic field through the Primakoff effect with a two-photon vertex \cite{sikivie1983experimental}. This coupling gives rise to the mixing of ALPs with photons \cite{raffelt1988mixing}, which leads to the conversion between ALPs and photons. This mechanism serves as the basis to search for ALPs, in particular, it has been put forward to explain a number of astrophysical phenomena, or to constrain ALP properties using observations. For example, over the last few years, it has been realized that this phenomenon would allow searches for the ALPs in the observations of distant active galactic nuclei (AGNs) in radio galaxies \cite{bassan2010axion, horns2012probing}. Since photons emitted by these sources can mix with ALPs during their propagation in the presence of an external magnetic field and this might reduce photon absorption caused by extragalactic background light \cite{harris2014photon}. This scenario might lead to a suitable explanation for the unexpected behavior for the spectra of several AGNs \cite{mena2013hints}. Furthermore, because only photons with polarization in the direction of the magnetic field can couple to ALPs, this coupling can lead to change in the polarization state of photons. This effect can also be useful to search for ALPs in the environment of AGNs by looking for changes in the linear degree of polarization from the values predicted by the synchrotron self-Compton model of gamma ray emission.
 
Many galaxy clusters show a soft X-ray excess in their spectra below $1 \textup{--} 2$ keV on top of the extrapolated high-energy power \cite{bonamente2002soft, durret2008soft}. This soft excess was first discovered in 1996 from the Coma and Virgo clusters, before being subsequently observed in many other clusters \cite{lieu1996discovery, lieu1996diffuse, bowyer1996extreme}. However, the nature of this component is still unclear. There are two astrophysical explanations for this soft X-ray excess phenomenon, for review see \cite{durret2008soft, angus2014soft}. The first is due to emission from a warm $T \approx 0.1$ keV gas. The second is based on inverse-Compton scattering of $\gamma \sim 300 \textup{--} 600$ non-thermal electrons on the Cosmic Microwave Background (CMB). However, both of these two explanations face difficulties to clarify the origin of the soft X-ray excess \cite{angus2014soft}. Recent work \cite{conlon2013excess} proposed that this soft excess is produced by the conversion of a primordial Cosmic ALP Background (CAB) with $0.1 \textup{--} 1$  keV energies into photons in the magnetic field of galaxy clusters. The existence of such a background of highly relativistic ALPs is theoretically well-motivated in models of the early universe arising from compactifications of string theory to four dimensions. Also, the existence of this CAB can be indirectly probed through its contribution to dark radiation, but this is beyond the scope of this work.
 
The main aim of this work is to test the CAB model that is put forward to explain the soft X-ray excess in the Coma cluster based on the conversion between CAB ALPs and photons in the presence of an external magnetic field using the M87 jet environment. We aim as well to demonstrate the potential of the AGNs jet environment to probe low-mass ALP models and to potentially constrain the model proposed to better explain the soft X-ray excess in the Coma cluster. We find that the overall X-ray emission for the M87 AGN will not be overproduced by the ALP-photon conversion model with very low ALP mass and very small ALP-photon coupling if a CAB is assumed and the misalignment angle between the line of sight and the AGN jet direction is not equal to zero. Our results suggest new constraints on the value of the ALP-photon coupling lower than the current limits used to explain the Coma cluster soft X-ray excess. This casts doubts upon the CAB model used in \cite{angus2014soft}.
 
The structure of this paper is as follows. In section \ref{sec.2}, we review the theoretical model that describes the ALP-photon mixing phenomenon. In section \ref{sec.3}, we briefly discuss the motivation for the existence of the CAB. Then in section \ref{sec.4}, we check whether the soft X-ray excess in the environment of the M87 AGN jet can be explained due to CAB ALPs conversion into photons in the jet magnetic field. In section \ref{sec.5}, the results of a numerical simulation of the ALP-photon coupling model are discussed and compared with observed soft excess luminosities in observations. Finally, our conclusion is provided in section \ref{sec.6}.

\section{ALP-photon coupling model} \label{sec.2}

We first outline the theory of the conversion between ALPs and photons in an external magnetic field within the environment of the jets of AGNs following \cite{mena2013hints, mena2011signatures}. In the presence of a background magnetic field, the coupling of an ALP with photon is described by the effective Lagrangian \cite{sikivie1983experimental, raffelt1988mixing, anselm1988experimental}

\begin{equation} \label{eq.1}
\mathrm{\ell}_{a\gamma} = - \frac{1}{4} g_{a\gamma} \mathrm{F}_{\mu \nu} \tilde{\mathrm{F}}^{\mu \nu} a = g_{a\gamma}\, \mathbf{E} \cdot \mathbf{B} \, a \:,
\end{equation}
where $g_{a\gamma}$ is the ALP-photon coupling parameter with dimension of inverse energy, $\mathrm{F}_{\mu \nu}$ and $\tilde{\mathrm{F}}^{\mu \nu}$ represent the electromagnetic field tensor and its dual respectively, and $a$ donates the ALP field. While $\mathbf{E}$ and $\mathbf{B}$ are the electric and magnetic fields respectively. Then, we consider a monochromatic and linearly polarized ALP-photon beam of energy $\omega$ propagating along the $z$-direction in the presence of an external and homogeneous magnetic field. The equation of motion for the ALP-photon system can be described by the coupled Klein-Gordon and Maxwell equations arising from the Lagrangian equation (\ref{eq.1}). For very relativistic ALPs when $\omega \gg m_a$, the short-wavelength approximation can be applied successfully and accordingly the beam propagation can be described by the following Schr$\ddot{\text{o}}$dinger-like form \cite{raffelt1988mixing, bassan2010axion}
\begin{equation} \label{eq.2}
\left( i \dfrac{d}{dz} + \omega + \mathcal{M} \right)
  \left( \begin{matrix} A_{\perp}(z) \\ A{\parallel}(z) \\ a(z) \end{matrix} \right) =0 \: ,
\end{equation}
where $A_{\perp}$ and $A_{\parallel}$ are the photon linear polarization amplitudes along the $x$ and $y$ axis, respectively, and $a(z)$ donates the ALP amplitude. Here,  $\mathcal{M}$ represents the mixing matrix of the ALP field with the photon polarization components. Since only photons with polarization parallel to the magnetic field couple to ALP, so for simplicity, we restrict our attention to the case of magnetic field transverse $\mathbf{B}_T$ to the beam direction (\rmfamily{i.e.} in the x-y plane). Therefore, if we choose y-axis along $\mathbf{B}_T$, so that $\mathbf{B}_x$ vanishes and the mixing matrix can be written as
\begin{equation} \label{eq.3}
\mathcal{M} = \left( \begin{matrix} 
\Delta_{\perp} & 0 & 0 \\ 0 &  \Delta_{\parallel} & \Delta_{a\gamma} \\ 0 & \Delta_{a\gamma} & \Delta_{a} \end{matrix} \right) \: .
\end{equation}
As expressed in references \cite{bassan2010axion, mena2013hints, mena2011signatures}, the elements of $\mathcal{M}$ and their references values are given as below
\begin{align}  \label{eq.4}
\Delta_{\perp}  &\equiv 2\; \Delta_{\text{QED}} + \Delta_{\text{pl}} \:,  \nonumber \\[10pt]
\Delta_{\parallel}  &\equiv \frac{7}{2}\; \Delta_{\text{QED}} + \Delta_{\text{pl}} \:,  \nonumber \\
\Delta_{a\gamma} &\equiv \frac{1}{2} \; g_{a\gamma} B_T \simeq 1.50 \times 10^{-11} \; \left( \frac{g_{a\gamma}}{10^{-10}\; \text{GeV}^{-1}} \right) \left( \frac{B_T}{10^{6} \; \text{G}} \right) \; \text{cm}^{-1} \:,  \nonumber \\
\Delta_a &\equiv  -\frac{m_a^2}{2\omega} \simeq -2.53 \times10^{-13}\; \left( \frac{\omega}{\text{keV}} \right)^{-1} \left( \frac{m_a}{10^{-7}\; \text{eV}} \right)^{2} \; \text{cm}^{-1} \:.
\end{align}
The two terms $\Delta_{\text{QED}}$ and $\Delta_{\text{pl}}$, account for the QED vacuum polarization and the plasma effects and they are determined as
\begin{align} \label{eq.5}
\Delta_{\text{QED}}  &\equiv \frac{\alpha \omega}{45 \pi} \left( \frac{B_T}{B_{cr}} \right)^{2} \simeq 1.34 \times10^{-12}\; \left( \frac{\omega}{\text{keV}} \right) \left( \frac{B_T}{10^{6}\; \text{G}} \right)^{2} \; \text{cm}^{-1} \:,  \nonumber \\
\Delta_{\text{pl}} &\equiv - \frac{\omega^2_{pl}}{2\omega}  \simeq -3.49 \times 10^{-12}\; \left( \frac{\omega}{\text{keV}} \right)^{-1} \left( \frac{n_e}{10^{8} \; \text{cm}^{-3}} \right) \; \text{cm}^{-1} \:.
\end{align}
Here, $\alpha$ is the fine structure and $B_{\text{cr}}=4.414\; \text{G}$ is the critical magnetic field. In a plasma, the photons acquire an effective mass given in term of the plasma frequency  $\omega^2_{\text{pl}}=4\pi \alpha n_e / m_e$, where $n_e$ is the plasma electron density.

For a general case when the transverse magnetic field $\mathbf{B}_T$ makes an angle $\xi$, (where $0 \leq \xi \leq 2\pi$), with the $y$-axis in a fixed coordinate system. A rotation of the mixing matrix (equation (\ref{eq.3})) in the x-y plane and the evolution equation of the ALP-photon system can then  be written as
\begin{equation} \label{eq.6}
i \dfrac{d}{dz} \left( \begin{matrix} A_{\perp}(z) \\ A{\parallel}(z) \\ a(z) \end{matrix} \right) = -  \left( \begin{matrix} 
\Delta_{\perp} \cos^2 \xi + \Delta_{\parallel} \sin^2 \xi & \cos \xi \sin \xi (\Delta_{\parallel}+\Delta_{\perp}) & \Delta_{a\gamma} \sin \xi  \\ 
\cos \xi \sin \xi (\Delta_{\parallel}+\Delta_{\perp})  & \Delta_{\perp} \sin^2 \xi + \Delta_{\parallel} \cos^2 \xi  & \Delta_{a\gamma} \cos \xi  \\
 \Delta_{a\gamma} \sin \xi & \Delta_{a\gamma} \cos \xi & \Delta_{a} \end{matrix} \right)
  \left( \begin{matrix} A_{\perp}(z) \\ A{\parallel}(z) \\ a(z) \end{matrix} \right) \: ,
\end{equation}
As we can see from the model equations, the conversion proportionality between ALPs and photons is very sensitive to the transverse magnetic field $\mathbf{B}_T$ and the plasma electron density $n_e$ profiles. However, their configurations in the AGN jets are not fully clear yet. In our work here, we adopted the following $\mathbf{B}_T$ and $n_e$ profiles
\begin{equation} \label{eq.7}
B_T(r,R) =J_s(r) \cdot  B_{\ast} \left( \frac{R}{R_{\ast}} \right)^{-1} \text{G} \:, \quad \text{and} \quad
n_e(r,R) = J_s(r) \cdot n_{e,\ast} \left( \frac{R}{R_{\ast}}  \right)^{-s} \; \text{cm}^{-3} \:.
\end{equation}
Where $r$ is the distance from the jet axis, $R$ is the distance along the jet axis from the central supermassive black hole, believed to be at the center of the AGN, and $R_{\ast}$ represents a normalization radius. The function $J_s(r)$ is the exponentially scaled modified Bessel function of the first kind of order zero \cite{abramowitz1965handbook, arfken1985mathematical}, used to scale the magnetic field and the electron density profiles with our choice here for the scale length to be three times the Schwarzschild radius of the central supermassive black hole. The normalization parameters $B_{\ast}$ and $n_{e,\ast}$ can be found in by fitting observational data for a given AGN with the suggested magnetic field and electron density profiles.  

Using the set of parameters discussed above, the evolution equations (\ref{eq.6}) can be numerically solved to find the two components of the photon linear politicization; $A_{\perp}$ and $A_{\parallel}$. If we consider the initial state is ALPs only, the initial condition is $(A_{\perp}, A_{\parallel}, a)^t = (0,0,1)$ at the distance $R = R_{\text{min}}$ where the ALPs enter the jet. Then the probability for an ALP to convert into a photon after traveling a certain distance in the magnetic field inside the AGN jet can be defined as \cite{raffelt1988mixing, mirizzi2008photon}
\begin{equation} \label{eq.9}
P_{a\rightarrow \gamma} (E) =  \vert A_{\parallel}(E) \vert^2 + \vert A_{\perp}(E) \vert^2  \:.
\end{equation}
Notice here that we follow the authors of reference \cite{mena2013hints} in assuming the jet field to be coherent along the studied scale and ALP-photon conversion probability is independent of the coherence length of the magnetic field. In addition, in \cite{marsh2017new} the authors find coherence lengths in M87 in excess of longest scales studied here. We will discuss later how the ALP-photon conversion probability is affected by the jet geometry and the direction of the beam propagation inside the jet. Moreover, the observations can be compared with the ALP-photon mixing model results by plotting the energy spectrum ($\nu F_{\nu} \equiv E^2 dN/dE$) as a function of the energy of the photons with $\omega \equiv E(1+z)$ where $z$ is the AGN redshift. The final photon spectrum is obtained by multiplying the photon production probability $P_{a\rightarrow \gamma} (E)$ with the CAB Band spectrum $S(E)$ \cite{band1993batse}
\begin{equation} \label{eq.10}
E^2 \dfrac{dN}{dE} = P_{a\rightarrow \gamma} (E) \cdot S(E) \:.
\end{equation}

Hence, the ALP-photon mixing for the AGN jet model includes six free parameters: the normalization for the magnetic field $B_{\ast}$, the normalization for the electron density $n_{e,\ast}$, the ALP mass $m_a$, the ALP-photon coupling parameter $g_{a\gamma}$, and two additional geometric parameters $\theta$ and $\phi$; we will discusses their role later in section \ref{sec.4}.

\section{Cosmic Axion Background} \label{sec.3}

In this section we follow reference \cite{conlon2013excess}, the authors motivate the existence of a homogeneous Cosmic Axion Background (CAB) with $0.1 \textup{--} 1$ keV energies arising via the decay of string theory moduli in the very early universe. The origin of this idea came from the four-dimensional effective theories arising from compactifications of string theory. A generic feature of these effective theories is the presence of massive scalar particles called moduli, coming from the compactification of massless fields. Theoretical arguments indicate that the existence of such moduli should be responsible for the reheating of the SM degrees of freedom. Regardless of the details of the inflation model, moduli are usually displaced from their final metastable minimum during inflation and begin to oscillate at the end of inflation. As they are characterized by their very weak, typically suppressed by Planck-mass interactions, the moduli are long-lived.  The oscillating moduli fields subsequently come to dominate the energy density of the universe which enters a modulus-dominated stage lasts until the moduli decay into visible and hidden sector matter and radiation, thus leading to reheating. The visible sector decays of the modulus rapidly thermalize and initiate the Hot Big Bang. The gravitational origin of the moduli implies that moduli can also decay to any hidden sector massless particles with extremely weak interactions, such as ALPs. Two-body decays of a modulus field $\Phi$ into ALPs are induced by the Lagrangian coupling $\frac{\Phi}{M_P} \partial_\mu a \partial^\mu a$, resulting in ALPs with an initial energy $E_a=m_\Phi/2$. 

Considering these ALPs arising from moduli decays at the time of reheating are weakly interacting, they do not thermalize and the vast majority of ALPs propagate freely to the present day. This implies that they would linger today forming a homogeneous and isotropic CAB with a non-thermal spectrum determined by the expansion of the universe during the time of moduli decay. Furthermore, because they are relativistic, they contribute to the dark radiation energy density of the universe, but this is beyond the scope of this work. For moduli masses $m_{\Phi} \approx$ GeV the present energy of these axions is $E_a \sim 0.1 \textup{--} 1$ keV. The suggestion being that the natural energy for such a background lies between $0.1$ and $1$ keV. Furthermore, in \cite{conlon2013cosmophenomenology} it was shown that such CAB would have a quasi-thermal energy spectrum with a peak dictated by the mass of the ALP. This CAB is also invoked in \cite{angus2014soft} to explain the soft X-ray excess on the periphery of the Coma cluster with an ALP mass of $1.1\times 10^{-13}$ eV and a coupling to the photon of $g_{a\gamma} = 2 \times 10^{-13}$ GeV$^{-1}$. 

In this work, we will assume, for convenience, that the CAB has a thermal spectrum with an average energy of $\langle E_a \rangle = 0.15$ keV. We then normalize the distribution to the typical example quoted in \cite{conlon2013cosmophenomenology}. We use the thermal distribution as an approximation, as the exact shape of the distribution will not substantially affect the conclusions we draw. We can then determine the fraction of CAB ALPs converted into photons within the environment of the M87 AGN jet and use this to determine a resulting photon flux. This flux can be compared to X-ray measurements to see if such environments can constrain low-mass ALPs and put limits on the ALP explanation of the X-ray excess.

\section{The soft X-ray excess CAB in the environment of M87 AGN jet} \label{sec.4}

It is established that most of the baryonic mass of the galaxy clusters are composed of a hot ionized intracluster medium (ICM) with temperatures about of $T \approx 10^8\; \text{K}$ (corresponding to $\omega \approx 7 \; \text{keV}$) and number densities in the range of  $n \sim 10^{-1} \textup{--} 10^{-3} \; \text{cm}^3$. The ICM generates diffuse X-ray emission through thermal bremsstrahlung. A thermal bremsstrahlung spectrum gives a good approximation of a constant emissivity per unit energy at low energies. However, observations of many galaxy clusters show a soft X-ray excess in their spectra at low energies around $1 \textup{--} 2$ keV, which is above that from the hot ICM, and the origin of this component is still unclear. In this work, we have adopted the scenario that this soft excess is produced by the conversion of a primordial Cosmic ALP Background (CAB) into photons in the cluster magnetic field. The central M87 AGN of the Virgo cluster is the best characterized AGN in the literature due to its proximity \cite{macchetto1997supermassive, gebhardt2009black, event2019first}. In this respective, we numerically solve the ALP-photon mixing model described in section \ref{sec.2} and use it to study the photon production probability from CAB ALPs in the environment of M87 AGN jet. Then, we test the model to reproduce the X-ray emission for the M87 AGN from the ALP-photon conversion with very low ALP mass and very small ALP-photon coupling.

The M87 AGN is a radio galaxy at a luminosity distance of $16.7 \pm 0.2$ Mpc \cite{mei2007acs} and a redshift of z = 0.00436. Based on its radio images and the modeling of its interaction with the surrounding environment, it is suggested that the M87 jet is misaligned with respect to the line of sight \cite{biretta1999hubble, biretta1995detection}.  Therefore, we consider the situation when there is a misalignment between the ALP-photon beam propagation direction and the AGN jet direction. Accordingly, we have to take into account the geometry of the jet of the AGN and the direction of the ALP-photon beam propagation. In this case, two more parameters may play an important role in the study of the ALP-photon conversion probability are the misalignment angle $\theta$ between the jet direction and the line of sight and the AGN jet opening angle $\phi$ which define the jet geometry. When $\theta=0$, the ALP-photon beam propagate parallel to the R-direction and we expect that the ALP-photon conversion probability is not affected by the jet geometry represented by the jet opening angle $\phi$. However, when the ALP-photon beam crosses the jet diagonally making a misalignment with the R-direction, the ALP-photon conversion probability may be affected by the misalignment angle $\theta$ as well as the jet opening angle $\phi$. The opening angle $\phi$ for the jet of the M87 AGN near the base is less than about 5 degrees \cite{biretta1995detection}, and the misalignment angle $\theta$ is less than about 19 degrees \cite{walker2008vlba, hada2017structure}.

In this study, we make our choices such that the magnetic field and the electron density profiles used for M87 AGN are consistent with the obtained values in \cite{park2019faraday}. In this perspective, we use an electron profile density profile $n_e \propto R^{-1}$ with $s=1$ in equation (\ref{eq.7}). For our case of varying $\mathbf{B}_T$ and $n_e$ with $R$ as in equation (\ref{eq.7}), transition of ALPs to photons take place over different distances,  $R \sim 10^{16} \textup{--} 10^{17}\; \text{cm}$, with normalization radius $R_{\ast} = 6 \times 10^{20}\; \text{cm}$. The environmental parameters are taken as $B_{\ast}= 1.4 \times 10^{-3}$ G and $n_{e,\ast}=0.3$ cm$^{-3}$ at the distance $R_{\ast}$. Note that as $n_e$ values are only available at larger distances from the base \cite{park2019faraday}, we assume that we can extrapolate its values down to small radii. In addition, we set the ALP mass to be $1.1\times 10^{-13}$ eV and we start with ALP-photon coupling around $g_{a\gamma}  \sim 2 \times 10^{-13}$ GeV$^{-1}$ in agreement with the models derived in \cite{angus2014soft} to explain the soft X-ray excess on the periphery of the Coma cluster. 

It is worth noting that, in a more general scenario, the actual numerical value of $m_a$ would have little influence on the constraints on the ALP-photon coupling. However, the presence of a CAB is what we are attempting to constrain. In particular, the CAB we are addressing is the one proposed to account for the soft X-ray excess in the Coma cluster and here $m_a$ influences the peak-energy of ALP distribution according to \cite{angus2014soft}, so we quote the value they provide as the best fit. Some models include \cite{turner1991inflationary, kawasaki2013axions, hertzberg2017correlation, cheung2012cosmological, khlopov1999nonlinear} predict the mass scale to be larger than the one we use here, however, those models are interested in studying various other ideas. This indeed seems to be an important subject to take into account for future work.

\section{Results and discussion} \label{sec.5}

\begin{figure}[tbp]
\centering
\includegraphics[width=.495\textwidth]{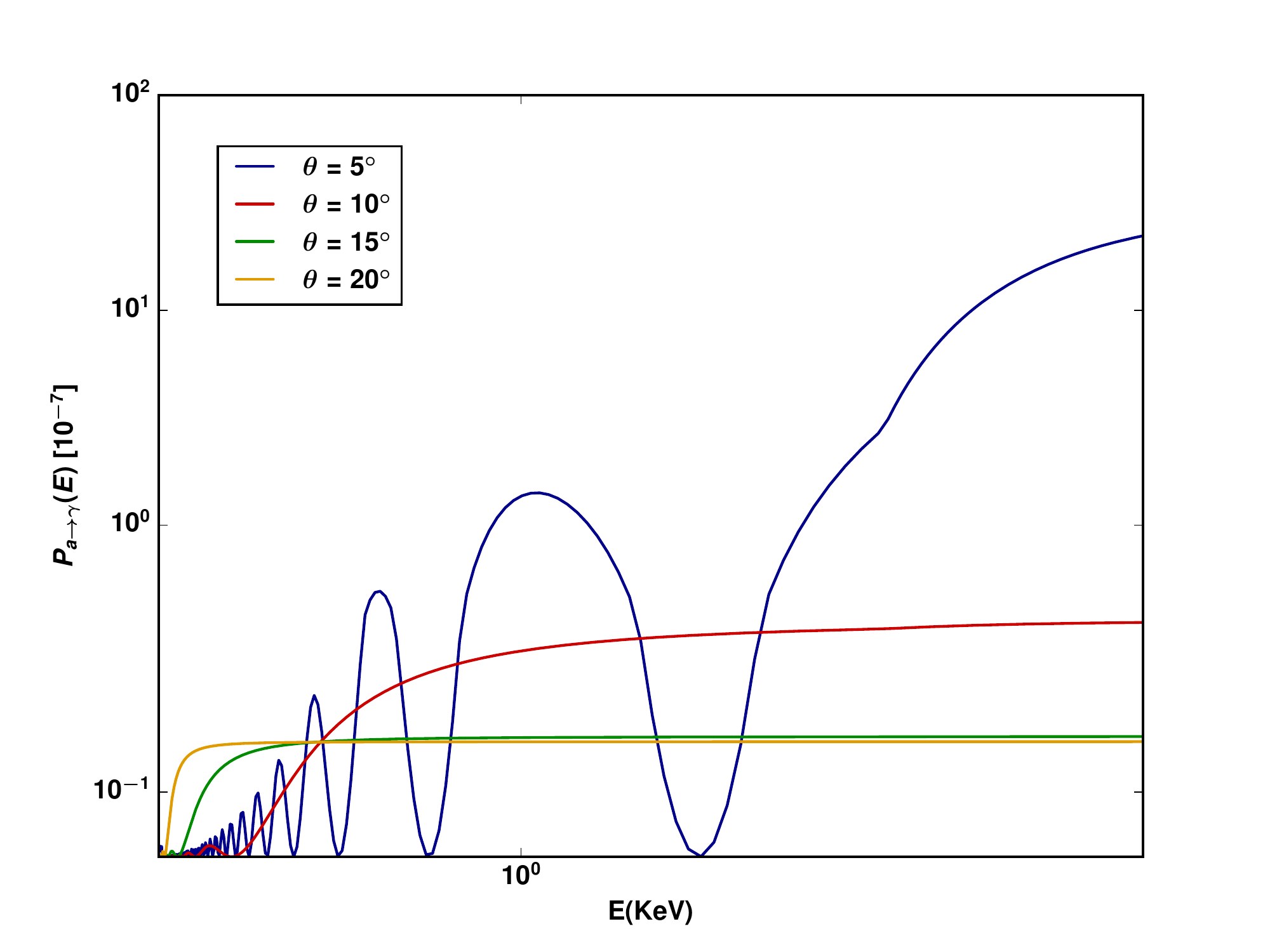}
\hfill
\includegraphics[width=.495\textwidth]{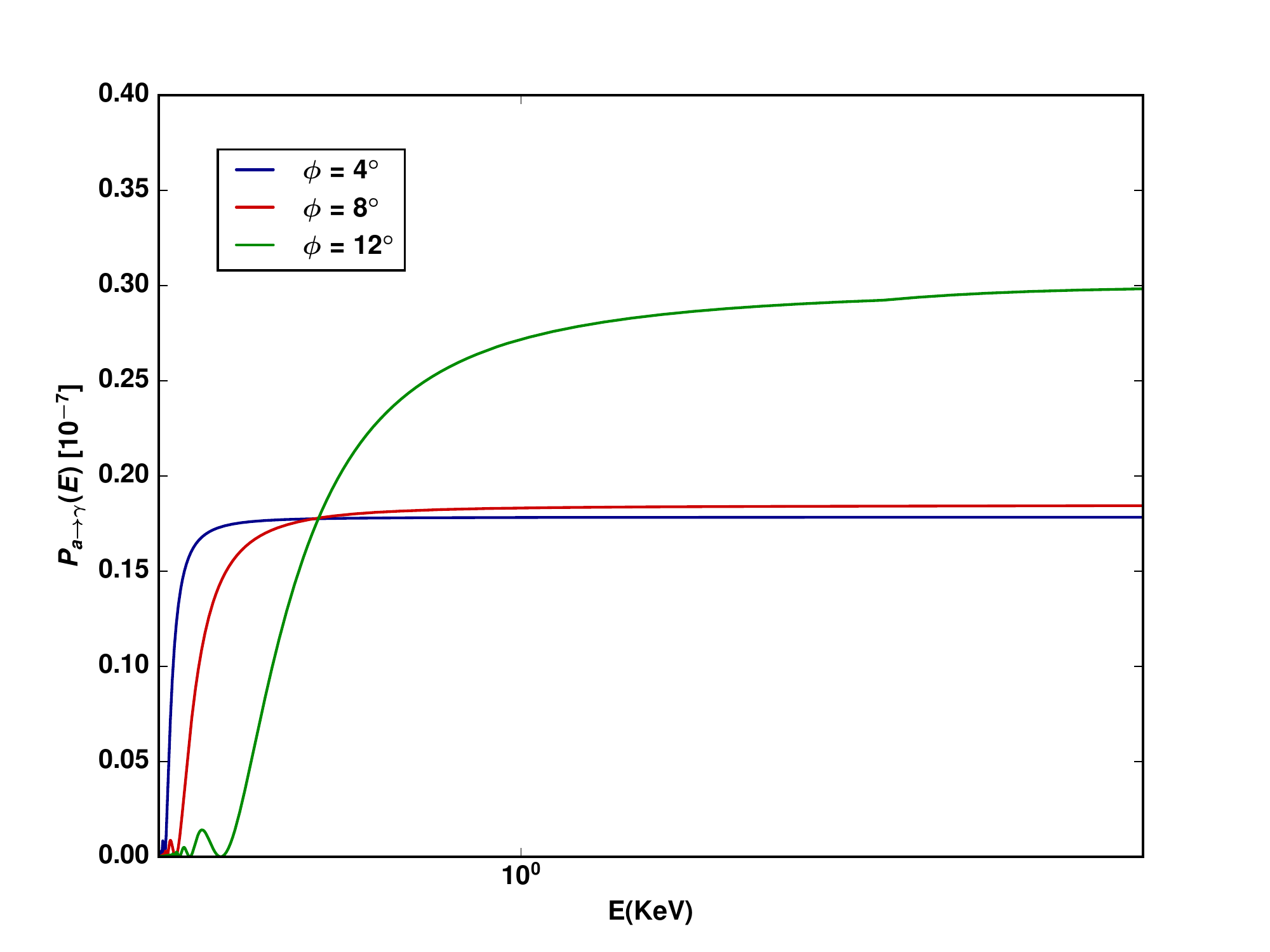}
\caption{Plot of the ALP-photon conversion probability $P_{a \rightarrow \gamma} (E)$. Left: The different curves correspond to $\theta = 5^\circ, 10^\circ, 15^\circ,$ and $20^\circ$ at fixed $\phi=4^\circ$. Right: The different curves correspond to $\phi = 4^\circ, 8^\circ,$ and $12^\circ$ at fixed $\theta=20^\circ$.}
\label{fig.1}
\end{figure}

In this section, we discuss the results of the numerical simulation of the conversion probabilities and present a description of the predictions of the scenario of CAB ALPs conversion to photons for the M87 AGN soft X-ray excess. To obtain our results, we apply the ALP-photon mixing model to study the probability of CAB ALPs to convert to photons in the intergalactic magnetic field on the jet of M87 AGN with the initial state of ALPs only at $R_{\text{min}}= 10^{16}\; \text{cm}$. Figure \ref{fig.1} shows the ALP-photon conversion probability $P_{a \rightarrow \gamma} (E)$ as a function of energy for different values of the misalignment angle $\theta$ and the jet opening angle $\phi$. The different curves on the left panel correspond to $\theta = 5^\circ, 10^\circ, 15^\circ,$ and $20^\circ$ at fixed $\phi=4^\circ$, while the different curves on the right panel correspond to $\phi = 4^\circ, 8^\circ,$ and $12^\circ$ at fixed $\theta=20^\circ$. It seems to be clear from the two graphs that the maximum conversion probability occurs when the misalignment angle $\theta$ is more close to the opening angle of $\phi$. This might be explained due to the relation between the ALP-photon beam direction and jet geometry. For the beam to cross the jet diagonally from one side to another making an arbitrary angle with the magnetic field between zero and $\pi$, the condition for the beam to make the longest path is that the misalignment angle $\theta$ to be very close (but not equal) to the opening angle $\phi$. We have to remark here that the longest path is less than the maximum distance $R_{\text{max}} = 10^{17}\; \text{cm}$ when there is no misalignment. The misalignment at a given opening angle controls the point at which the ALP-photon beam would leave the jet and therefore the total distance that the beam travels inside the jet. Since we selected an electron density profile $n_e \propto R^{-1}$, this defines the relationship between the misalignment and the electron density profile that sets the critical energy at which stronger mixing starts as shown by the plots where less misaligned cases have higher critical energies and thus less overlap with the CAB spectrum.

\begin{figure}[tbp]
\centering
\includegraphics[width=.495\textwidth]{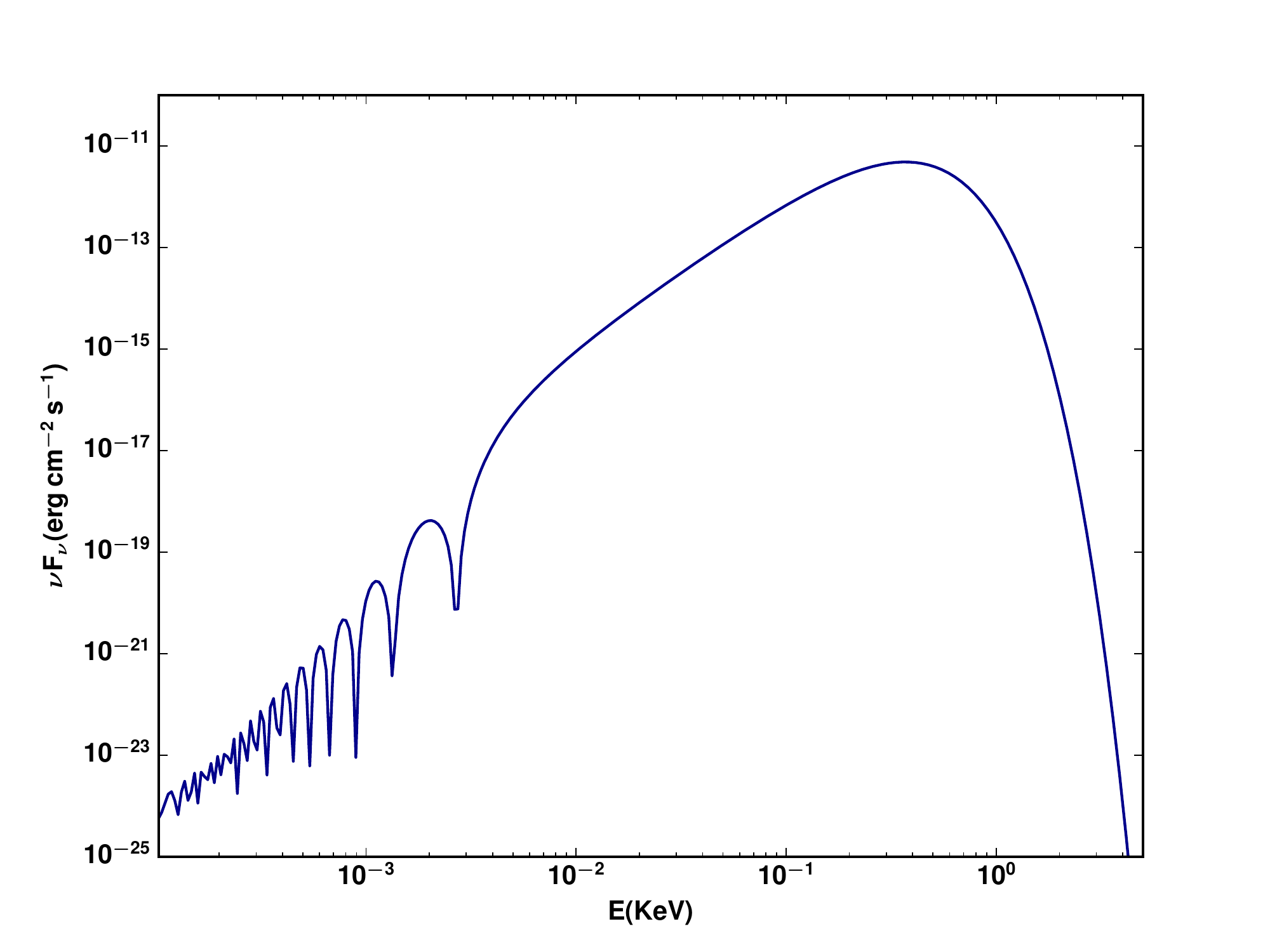}
\hfill
\includegraphics[width=.495\textwidth]{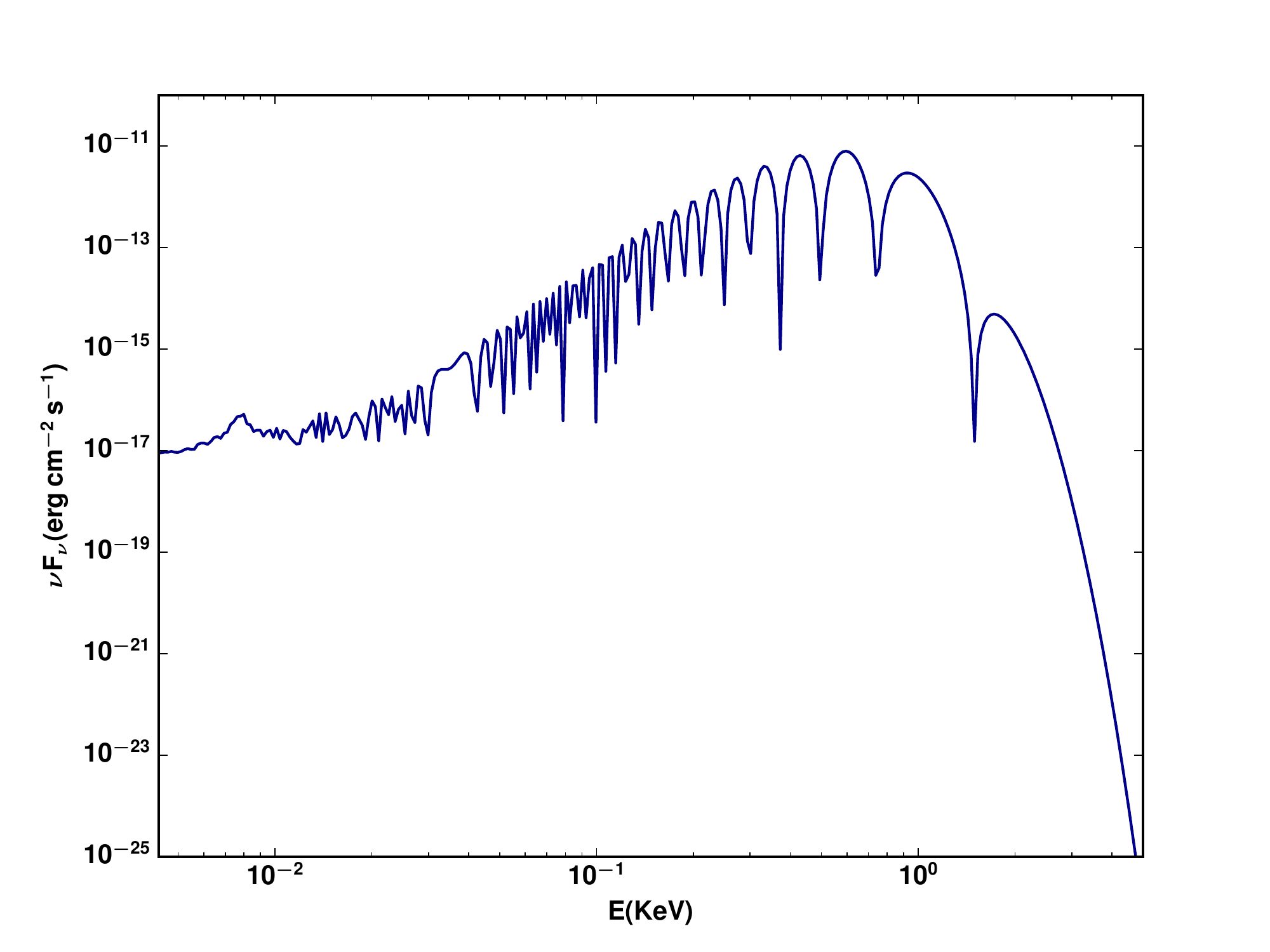}
\vfill
\includegraphics[width=.495\textwidth]{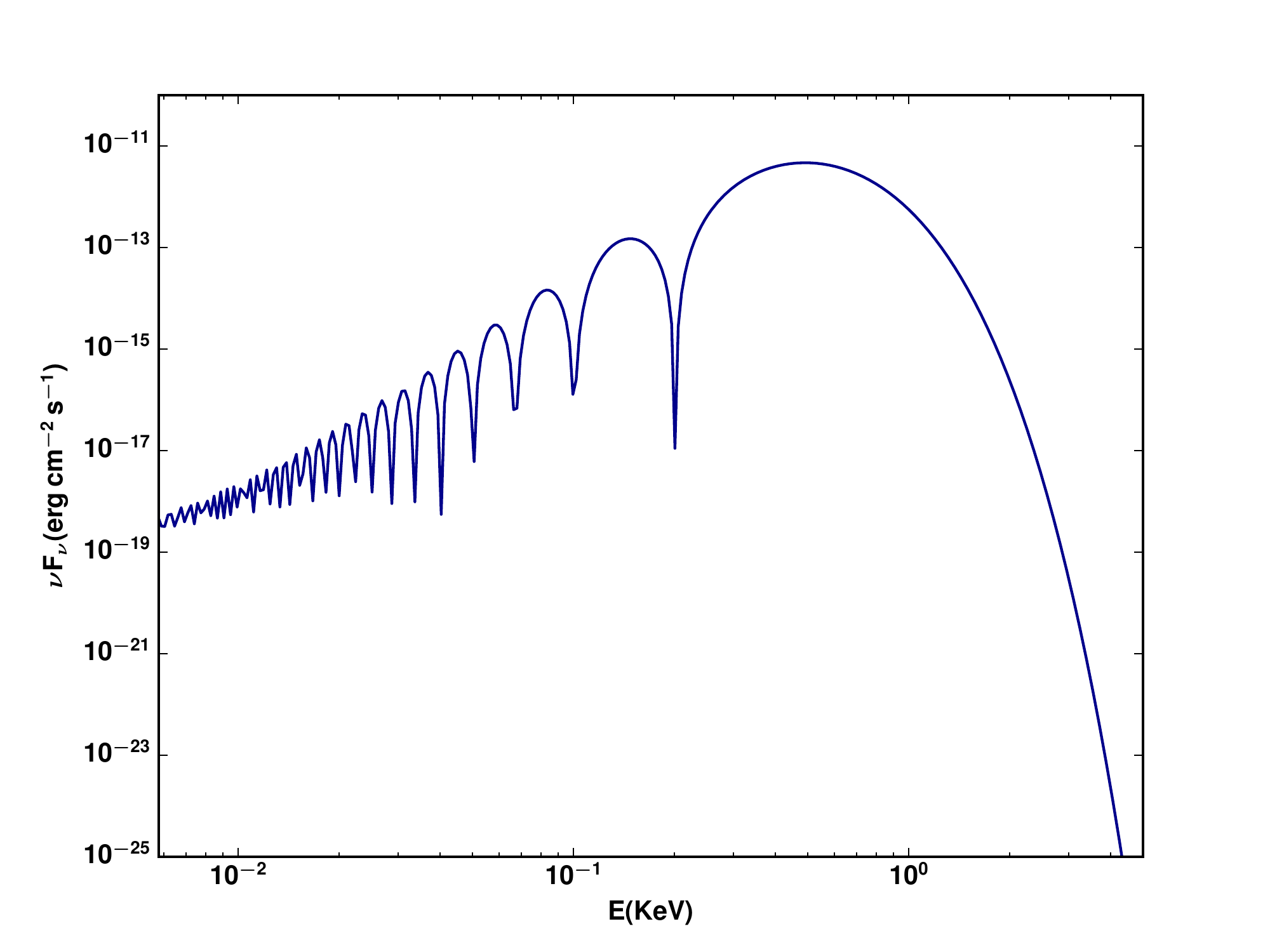}
\hfill
\includegraphics[width=.495\textwidth]{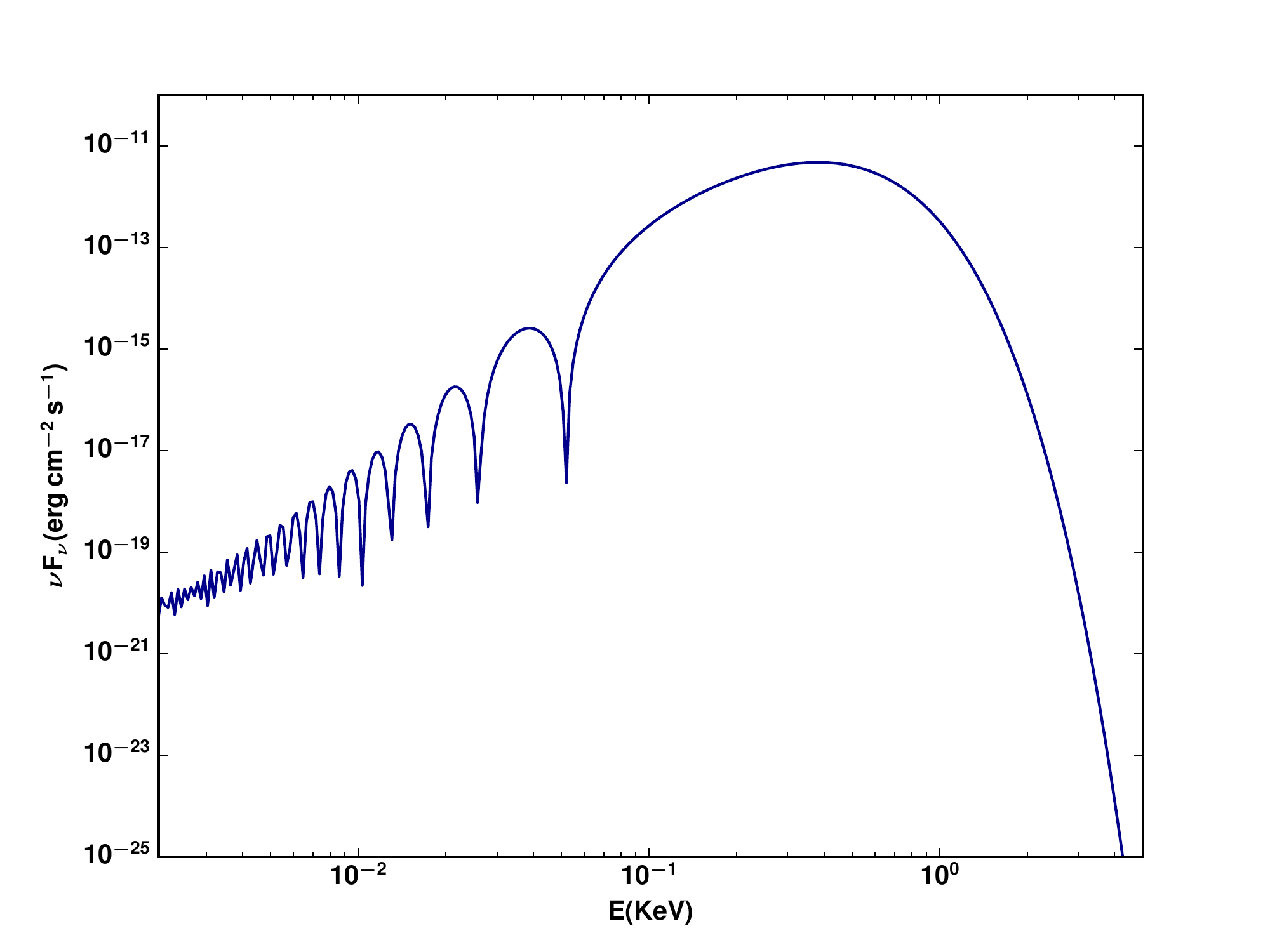}
\vfill
\includegraphics[width=.495\textwidth]{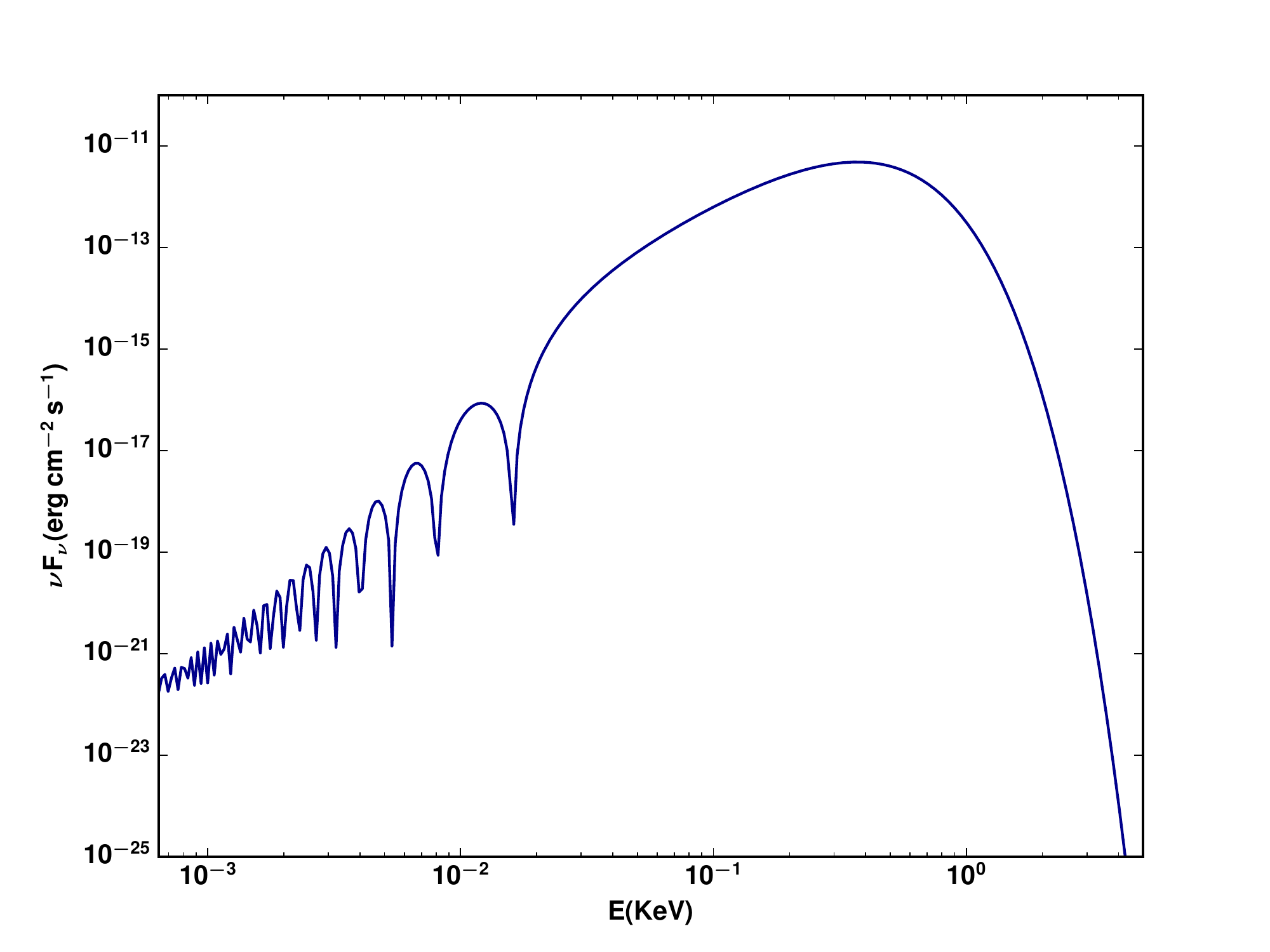}
\hfill
\includegraphics[width=.495\textwidth]{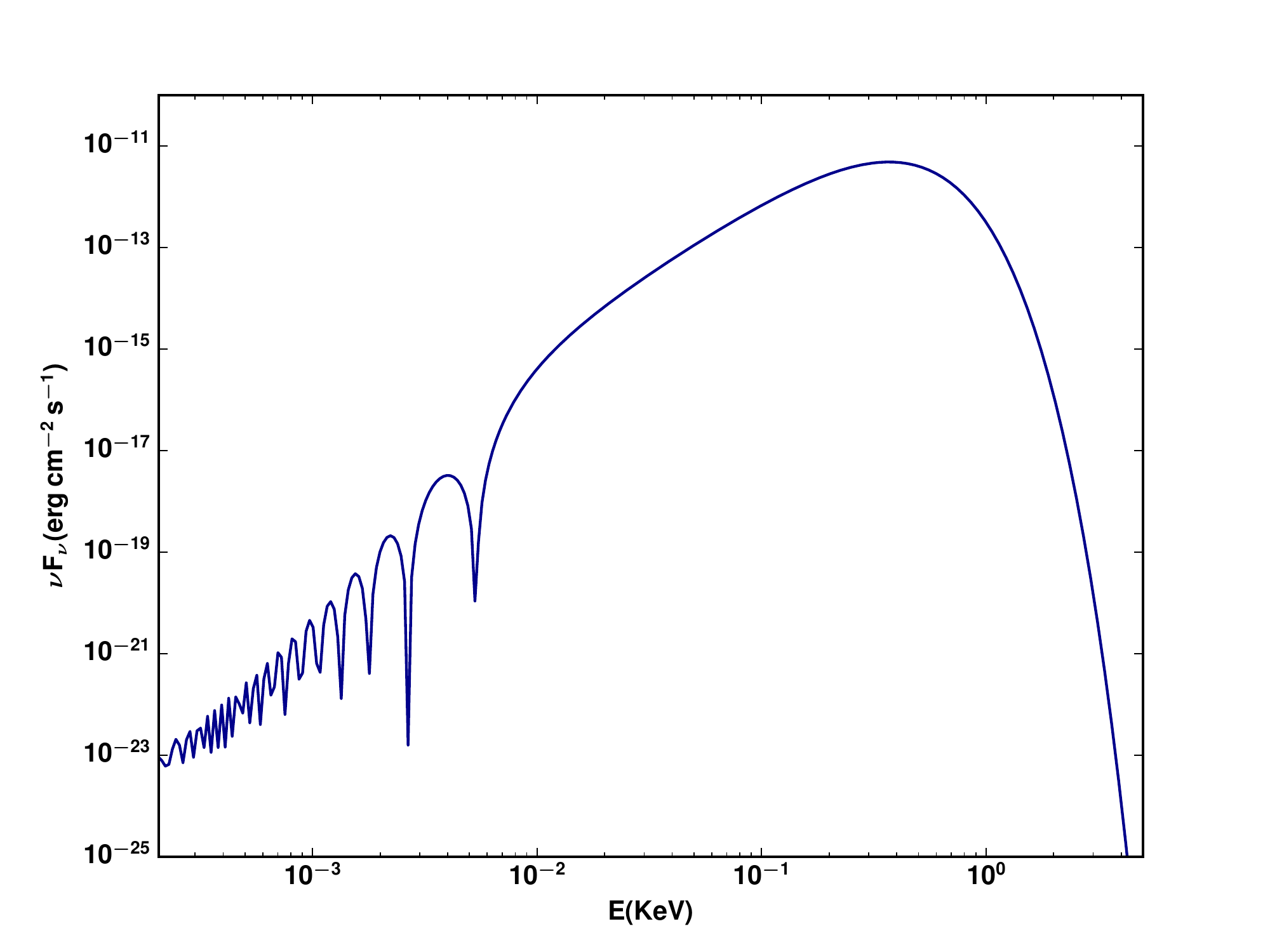}
\caption{The numerical simulation of the energy spectrum distributions from ALPs conversion to photons in the intergalactic magnetic field on the jet of M87 AGN at fixed opening angle $\phi =4^\circ$ and different values of the misalignment angle $\theta$: Top left, $\theta=0^\circ$. Top right, $\theta=5^\circ$. Middle left, $\theta=10^\circ$. Middle right, $\theta=15^\circ$. Bottom left, $\theta=20^\circ$. Bottom right, $\theta=25^\circ$.}
\label{fig.2}
\end{figure}

At this stage, we are ready to present the output of the model and check to put some new constants on the acceptable limit of the value of the ALP-photon coupling parameter $g_{a\gamma}$. Figure \ref{fig.2} shows our results for the energy spectrum distributions obtained from the numerical simulation for the ALPs conversion to photons in the intergalactic magnetic field of the jet of M87 AGN. For these plots, we kept using plasma electron density profile $n_e \propto R^{-1}$ as taking the parameter $s=1$ in equation (\ref{eq.7}) as well as using opening angle $\phi=4^\circ$ for the jet of M87 AGN. The different plots then represent the energy spectrum distributions at different values of the misalignment angle that the ALP-photon beam makes with the jet direction $\theta = 0^\circ, 5^\circ, 10^\circ, 15^\circ, 20^\circ,$ and $25^\circ$. For each case, we find the maximum value for the ALP to Photon coupling $g_{a\gamma}$ such that we do not exceed the observed flux $\sim 3.76 \times 10^{-12}\; \text{erg}\;\text{cm}^{-2}\;\text{s}^{-1}$ from the M87 AGN between $0.3$ and $8$ $\text{keV}$ \cite{m87chandra}. Table \ref{tab.1} shows six different cases of  $\theta = 0^\circ, 5^\circ, 10^\circ, 15^\circ, 20^\circ, 25^\circ$ at fixed $\phi=4^\circ$ and three different cases of $\phi = 4^\circ, 8^\circ, 12^\circ$ at fixed $\theta=20^\circ$ with the corresponding constraints the model put on the ALP to Photon coupling to produce the correct flux that is compatible with observations. The summary of the results presented in the table shows that the model put constraints on the value of ALP to photon coupling to be about $\sim 7.50 \times 10^{-15}  \textup{--} 6.56 \times 10^{-14} \; \text{GeV}^{-1}$ for ALP masses $m_a \lesssim  10^{-13} \; \text{eV}$ if there is a misalignment between the AGN jet direction and the line of sight less than 20 degrees.

It is important to note that the bounds on $g_{a\gamma}$ we derive in this paper are stronger than those found in \cite{marsh2017new}, which also simulates the effects of ALP-photon conversion in the environment of M87. The advantage in terms of constraints comes because we specifically study a CAB whereas the authors in \cite{marsh2017new} consider the loss of photons to ALP interactions in general, rather than the addition of photons from a cosmic ALP flux. The very large magnitude of the background ALP flux is what allows us to achieve such strong constraints. Additionally, we have to note that \cite{marsh2017new} derive more universal limits, as they do not require a CAB to exist for their limits to be valid.
\begin{table}[tbp]
\centering
\begin{tabular}{|c|c|c|c|}
\hline
$\:$ $\theta$ (${}^\circ$), $\phi= 4^\circ$ $\:$& $\qquad$ $g_{a\gamma}$  $(\text{GeV}^{-1})$ $\qquad$ &$\:$ $\phi$ (${}^\circ$), $\theta = 20^\circ$ $\:$ & $\qquad$ $g_{a\gamma}$  $(\text{GeV}^{-1})$ $\quad$ \\
\hline
$0$ &  $\lesssim 3.91 \times 10^{-13}$ &$4$ &  $\lesssim 6.56 \times 10^{-14}$\\
$5$ & $\lesssim 9.17 \times 10^{-15}$&$8$ &  $\lesssim 2.32 \times 10^{-14}$\\
$10$ & $\lesssim 7.50 \times 10^{-15}$&$12$ &  $\lesssim 7.99 \times 10^{-15}$\\
$15$ & $\lesssim 2.08 \times 10^{-14}$&&\\
$20$ & $\lesssim 6.56 \times 10^{-14}$&&\\
$25$ & $\lesssim 1.98 \times 10^{-13}$&&\\
\hline
\end{tabular}
\caption{The ALP to photon coupling $g_{a\gamma}$ corresponds to different values of the misalignment angle $\theta$ and the jet opening angle $\phi$ for the M87 AGN at which we produce the correct flux that is compatible with observations.}
\label{tab.1} 
\end{table}

\section{Conclusion} \label{sec.6}

In this work, we have reviewed the phenomenology of the conversion between very low-mass ALPs and photons in the presence of external magnetic fields. Then we tested the CAB model argued in \cite{angus2014soft} to explain the soft X-ray excess on the Coma cluster periphery. We did this by calculating the X-ray emission due to CAB propagation in the jet of the M87 AGN. It is evident in these results that the overall X-ray emission for the M87 AGN, between $0.3$ and $8$ keV, can be reproduced via the photons production from CAB ALPs with coupling $g_{a\gamma}$ in the range of $\sim 7.50 \times 10^{-15}  \textup{--} 6.56 \times 10^{-14} \; \text{GeV}^{-1}$ for ALP masses $m_a \lesssim 10^{-13} \; \text{eV}$ if there is a misalignment between the AGN jet direction and the line of sight less than 20 degrees, since the M87 jet has been found to be misaligned by less than 19 degrees in \cite{walker2008vlba, hada2017structure}. These values are up to an order of magnitude smaller than the current best fit value on the ALP-photon coupling $g_{a\gamma} \sim 2 \times 10^{-13} \; \text{GeV}^{-1}$ obtained in soft X-ray excess CAB model for the Coma cluster \cite{angus2014soft}. This casts doubt on the current limits of the largest allowed value of the ALP-photon coupling as a universal background would need to be consistent with observations of any given environment. Thus, our results exclude the current best fit value on the ALP-photon coupling for the Coma soft X-ray excess CAB model from \cite{angus2014soft} and instead place a new constraint that $g_{a\gamma} \lesssim 6.56 \times 10^{-14} \; \text{GeV}^{-1}$ when a CAB is considered.

\acknowledgments

This work is based on the research supported by the South African Research Chairs Initiative of the Department of Science and Technology and National Research Foundation of South Africa (Grant No 77948). A. Ayad acknowledges support from the Department of Science and Innovation/National Research Foundation (DSI/NRF) Square Kilometre Array (SKA) post-graduate bursary initiative under the same Grant. G. Beck acknowledges support from a National Research Foundation of South Africa Thuthuka (Grant No 117969). The authors would like also to offer special thanks to Prof. S. Colafrancesco, who, although no longer with us, continues to inspire by his example and dedication to the students he served over the course of his career. 

\bibliographystyle{unsrt}
\bibliography{references}

\end{document}